\documentclass[twocolumn]{IEEEtran}
\usepackage{setspace}
\usepackage{graphics}
\usepackage{colortbl}
\usepackage{dblfloatfix}
\usepackage[T1]{fontenc}
\usepackage{siunitx}
\usepackage{pgfplots}
\usepackage{grffile}
\pgfplotsset{compat=newest}
\usetikzlibrary{plotmarks}
\usetikzlibrary{arrows.meta}
\usepgfplotslibrary{patchplots}
\usepackage{amsmath}

\usepackage{hyperref}
\hypersetup{
    colorlinks=true,
    linkcolor=blue,
    filecolor=magenta,      
    urlcolor=blue,
}

\usetikzlibrary{plotmarks}
\usetikzlibrary{backgrounds}
\usetikzlibrary{fit}
\usetikzlibrary{positioning}
\usetikzlibrary{shapes}
\usepgfplotslibrary{patchplots}
\usepackage{grffile}
\usepackage{amsmath}
\usepackage{balance}
\usepackage{array, makecell, booktabs, rotating}
\usepackage{enumitem}
\usetikzlibrary{calc}
\usepackage{xcolor}
\pgfplotsset{plot coordinates/math parser=false}
\newlength\figureheight
\newlength\figurewidth
  
\usepackage{breqn}
\usepackage{flafter} 
\usepackage{placeins}

\usepackage{siunitx}
\usepackage{graphicx}
\usepackage[english]{babel}
\usepackage{eqnarray}

\usepackage{multirow}
\usepackage[noadjust]{cite} 
\usepackage{arydshln}
\usepackage{longtable}
\usepackage{multirow}
\usepackage{array}

\usepackage{url}
\definecolor{applegreen}{rgb}{0.7, 1, 0.0}

\newcolumntype{M}[1]{>{\centering\arraybackslash}m{#1}}
\newcolumntype{N}{@{}m{0pt}@{}}

\usepackage{amssymb}
\usepackage{color}
\usepackage{textcomp}
\usepackage{diagbox}

\usepackage{subfigure}
\usepackage[percent]{overpic}
\usepackage{float}
\usepackage{algorithm}
\usepackage{algpseudocode}
\newcommand*{\rom}[1]{\expandafter\@slowromancap\romannumeral #1@}

\makeatletter
\renewcommand{\ALG@beginalgorithmic}
\makeatother

\usepackage{graphicx}
\usepackage[english]{babel}
\usepackage{eqnarray,amsmath}
\usepackage{multirow}
\usepackage[noadjust]{cite} 
\usepackage{siunitx}
\usepackage{tikz}
\usepackage[cuteinductors,smartlabels]{circuitikz}

\usepackage{amssymb}
\usepackage{color}
\usepackage{textcomp}
\usepackage{comment}
\usepackage[normalem]{ulem} 
\usepackage{palatino}

\newcommand{\ihab}[1]{{\textcolor{black}{#1}}}
\newcommand{\kasim}[1]{{\textcolor{black}{#1}}}
\newcommand{\parisa}[1]{{\textcolor{black}{#1}}}
\newcommand{\intissar}[1]{{\textcolor{black}{#1}}}

\newcommand{\kasimm}[1]{{\textcolor{black}{#1}}}
\newcommand{\parisaa}[1]{{\textcolor{black}{#1}}}
\begin{document}

\title{An Artificial Neural Network-Based Model Predictive Control for Three-phase Flying Capacitor Multi-Level Inverter}

\author{Abualkasim Bakeer$^{1, \ast}$\!, Ihab S. Mohamed$^{2, \ast}$\!, Parisa Boodaghi Malidarreh$^{3, \ast}$\!, Intissar Hattabi$^{4}$, Lantao Liu$^{2}$
\thanks{$^{1}$ Abualkasim Bakeer is with the Department of Electrical Engineering, Faculty of Engineering, Aswan University, Aswan 81542, Egypt; {\tt\small {abualkasim.bakeer}@aswu.edu.eg}}
\thanks{$^{2}$ Ihab S. Mohamed and Lantao Liu are with the Luddy School of Informatics, Computing, and Engineering, Indiana University, Bloomington, IN 47408 USA, {\tt\small \{mohamedi, lantao\}@iu.edu}}
\thanks{$^{3}$ Parisa Boodaghi Malidarreh is with the Department of Power Electronics and Electrical Machines, Iran University of Science and Technology (IUST), Iran, {\tt\small pboodaghi}@gmail.com}
\thanks{$^{4}$ Intissar Hattabi$^{4}$ is with the Department of Automatic and Electrical Engineering, Saad Dahlab University of Science and Technology, BLIDA, Algeria  {\tt\small hattabi\_intissar@univ-blida.dz}}
\thanks{$^{\ast}$ Abualkasim Bakeer, Ihab S. Mohamed, and Parisa Boodaghi Malidarreh contributed equally to this work as the first author.}
}%
\maketitle

\begin{abstract}
Model predictive control \textit{(MPC)} has been used widely in power electronics due to its simple concept, fast dynamic response, and good reference tracking. However, it suffers from parametric uncertainties, since it directly relies on the mathematical model of the system  to predict the optimal switching states to \textcolor{black}{be used} at the next sampling time. As a result, uncertain parameters lead to an ill-designed \textit{MPC}. 
\textcolor{black}{Thus}, this paper offers a model-free control strategy on the basis of artificial neural networks \textit{(ANNs)}, for mitigating the effects of parameter mismatching while having a little negative impact on the inverter's performance.
This method includes two related stages. First, \textit{MPC} is used as an expert to control the \textcolor{black}{studied converter} in order to \textcolor{black}{provide} a dataset, while, in the second stage, the obtained dataset is utilized to train the proposed \textit{ANN}. 
The case study herein is based on a four-level three-cell flying capacitor inverter. In this study, MATLAB/Simulink is used to simulate the performance of the proposed method, taking into account various operating conditions. Afterward, the simulation results are reported in comparison with the conventional \textit{MPC} scheme, demonstrating \textcolor{black}{the superior performance of the proposed control strategy in terms of robustness against parameters mismatch and low total harmonic distortion (\textit{THD}), especially when changes occur in the system parameters, compared to the \textcolor{black}{conventional} \textit{MPC}. \kasim{Furthermore, the experimental validation of the proposed method is provided based on the Hardware-in-the-Loop (HIL) simulation using the C2000TM-microcontroller-LaunchPadXL TMS320F28379D kit, demonstrating the applicability of the \textit{ANN}-based control strategy to be implemented on a DSP controller.
}}
\end{abstract}


\section{Introduction}
\label{sec:introduction}
\IEEEPARstart{P}{ower} converters have been used in a wide range of applications in recent decades due to their improved performance and efficiency \cite{cortes2009model}. In this context, substantial attention has been paid to multilevel inverters \textit{(MLIs)} that introduced with certain advantages, such as: (i) reducing the total harmonic distortion, (ii) reducing common-mode voltage, and (iii) reducing the $dv/dt$ (that is, the rate of voltage change over time in switching instant) stress that leads to a reduction in electromagnetic emissions. There are a number of well-established and conventional \textit{MLI} topologies that have been implemented and refined over time, such as: flying capacitor \textit{(FC)}, cascaded H-Bridge \textit{(CHB)}, neutral point clamped \textit{(NPC)}, as well as innovative topologies that combine different topologies to improve \textit{MLIs} performance \cite{habetler2002design}. As a result, different control methods have been investigated including various linear methods such as proportional-integral \textit{(PI)} with pulse width modulation \textit{(PWM)} or space vector modulation \textit{(SVM)} \cite{jung2004optimal,mohamed2013model}, non-linear methods with Hysteresis \cite{mohamed2013classical}, 
 and other non-classical (i.e., modern) control methods  such as different techniques of predictive control \cite{rodriguez2012predictive, mohamed2013classical}.
However, controlling the \textit{MLIs} with the high number of switches brings about dire problems in controlling algorithms and makes them more complex. Moreover, voltage imbalance among the \textit{MLI} capacitors becomes an important concern in designing efficient control algorithms \cite{kouro2010recent,kouro2012power,rodriguez2007multilevel,franquelo2008age}. 

Model predictive control, among several control methods described, has appealing aspects that demonstrate its proper functioning. It provides a fast transient response, straightforward implementation, and non-linear constraints consideration \cite{rodriguez2012state,cortes2007three,kwak2013switching, mohamed2015improved}. Furthermore, the development of efficient microprocessors accounts for applying the \textit{MPC} schemes widely in machine drive, grid-connected converters, and power supplies \cite{defay2010direct,papafotiou2008model,song2013predictive,  mohamed2016implementation}.
By explicitly utilizing the model of the system to be controlled, \textit{MPC} selects, at each time step, the optimal switching state that minimizes a pre-defined cost function. 
However, it has a high \textit{online} computational burden especially when multiple time-horizons are considered, as: (i) the cost function should be calculated for different possible switching states; (ii) the optimization yields the optimal control sequence (i.e., optimal switching states); (iii) the first optimal switching state is finally applied to the system in the next sampling time.
 Sequentially, it can be quite difficult to employ the \textit{MPC} to control \textit{MLIs} with numerous levels \cite{findeisen2007assessment,nauman2016efficient,norambuena2017finite, wang2021model}. 
In addition, as previously mentioned, \textit{MPC} directly uses the mathematical model of the system, and thus the prediction accuracy is sensitive to the change in system parameters. 
As a result, it is susceptible to perform incorrectly as a result of parameter variations that may occur in the real system due to degradation and temperature effects \cite{kwak2014predictive}.

To be more specific, some non-modeled variables, such as temperature and magnetic saturation, have impacts on the real system's parameters that are so-called modeling errors or errors in measurements; these effects can deteriorate the control performance. 
Therefore, various control techniques have been proposed for tackling and analyzing the effects of parametric uncertainties \cite {martin2017sensitivity,young2016analysis}.  
On the other hand, since \textit{MPC} possesses a non-linear structure, these studies were carried out empirically by studying the behavior of
\textit{MPC} under different uncertainty conditions \cite{young2016analysis}, the obtained results led to adjust the model parameters such as inductance to mitigate the effects of the parameter mismatch. Moreover, the negative effects of mismatches in load resistance were ignored \cite{kwak2014predictive,bogado2013sensitivity}. 
For instance, authors in \cite {bogado2013sensitivity} examined the impact of electrical parameters' variation in six-phase induction motor drive. It has been shown that the parameter of inductance can affect the performance of \textit{MPC}, especially the stator and rotor leakage. 
While, in \cite {kwak2014predictive}, the predictive control-based direct power control with an adaptive parameter identification technique is proposed to cope with the parameter uncertainty problem. 
In this method, the input resistance and inductance of the converter are estimated using the input current and voltage at each sampling time, without having extra sensors. Thus, the converter parameters are updated online to curb the effects of parameter uncertainty. 

Furthermore, the use of data-based control approaches, such as \textit{ANN}-based control methods, has increased in recent research, particularly in power electronics and motor drives \cite{karanayil2011artificial, khan2021artificial}. Just to name a few, in \cite {mohamed2019neural}, an \textit{ANN}-based \textit{MPC} control strategy was introduced to improve the performance of a three-phase inverter with an output \textit{LC} filter, with comparison to the conventional \textit{MPC} scheme. \textcolor{black}{While, in \cite{wang2021model}, a general \textit{ANN-MPC} scheme is presented to tackle the computational challenge in adopting \textit{MPC} in highly complex power converters.} In \cite{liu2019finite}, the authors proposed a data-based technique for calculating the voltages of capacitors in order to eliminate the need for system sensors. While, an \textit{ANN}-based method is used in \cite{dragivcevic2018weighting} to regulate the weighting factors in the cost function of \textit{MPC}, which represents one of the research challenges in Model Predictive Control. Moreover, data-based control schemes are used in motor drives for estimating rotor speed, rotor flux, and torque in induction motors  \cite{wishart1995identification,lee2018performance,sun2012speed}. \kasim{In  \cite{bakeer2022efficient}, the authors applied \textit{ANN} to  calculate the dwell-time of the 3-phase \textit{2L-VSI}, which can improve the THD at a fixed-switching frequency operation.}

\kasim{This paper proposes an artificial neural network-based (\textit{ANN}-based) control strategy for a three-phase four-level flying capacitor inverter. 
\textit{MPC} has been used as a baseline control scheme for (i) collecting the data required for training process, considering different operating conditions, 
and (ii) assessing our proposed control strategy.
The main contributions
of the proposed \textit{ANN}-based control strategy can be summarized as
follows:
\begin{enumerate}
\item It is an \textit{End-to-End} (\textit{E2E}) control scheme that generates directly the optimal switching states of the inverter, without the need for either the mathematical model of the inverter or a pre-defined cost function to be minimized at each time step. 
\item It reduces the effects of the parameter mismatch problems as it offers a model-free control strategy based on \textit{ANNs}.
\item The proposed technique exhibits low \textit{THD} in the output current 
compared to the conventional \textit{MPC}.
\item \kasim{The HIL simulation is utilized to better assess the performance of the \textit{ANN}-based control method comapared with conventional \textit{MPC}. }
\item Moreover, the impact of the input features selection on the performance of the proposed control strategy is studied and evaluated. \ihab{Note that points 4) and 5) have not been addressed in the originally proposed method in \cite{mohamed2019neural}}. 
\item Finally, we provide an open repository of the dataset and simulation files to the community for further research activities\footnote{\url{https://github.com/IhabMohamed/ANN-FCMLI}}.
\end{enumerate}
}

This paper is organized as follows. In Section \ref{System Model}, the flying capacitor multilevel inverter \textit{(FCMLI)} and its mathematical model are briefly described, whereas the model predictive control strategy is explained in Section \ref{MPC}. The \textit{ANN}-based \textit{MPC} control strategy proposed in this work for the \textit{FCMLI} is presented in Section \ref{proposed-ANN-based}. Simulation results for both control strategies are given in Section \ref{simulations}. \kasim{In Section \ref{experimental result}, the HIL simulation is utilized to show the efficiency of the proposed \textit{ANN}-based control strategy in comparison with the \textit{MPC} scheme.} Finally, conclusions are given in Section \ref{conclusion}.
\begin{figure}[!ht]
\renewcommand{\figurename}{Fig.}
\begin{center}
\hspace*{-0.4cm}\includegraphics[scale=.67]{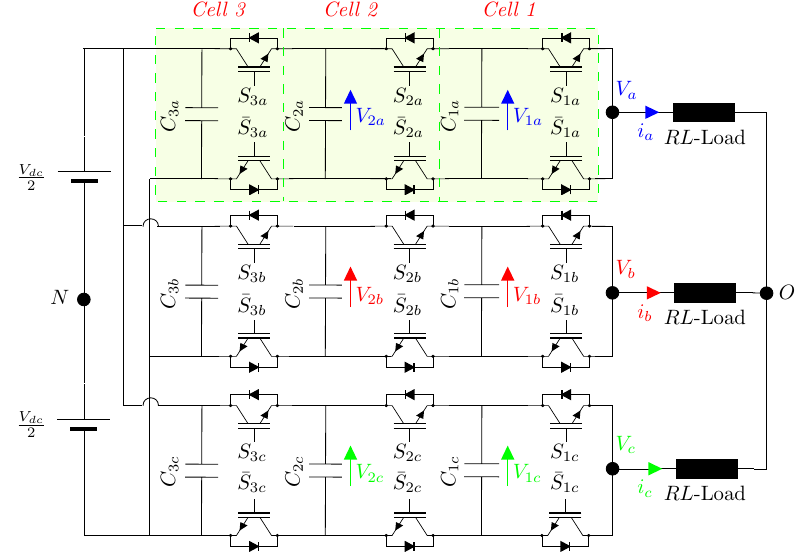}
\vspace*{-14pt}
\caption{Topology of a three-phase four-level flying capacitor inverter that is directly connected to an $RL$-load.}
\label{fig: Topology_system}
\end{center}
\end{figure}

\section{Flying Capacitor Multilevel Inverter}\label{System Model}
Flying capacitor multilevel inverter \textit{(FCMLI)} comprises series of connected cells which have a flying capacitor and two switches with antiparallel diodes. These switches are working in a complementary manner to avoid the short-circuit between the flying capacitor terminals. An $N$-cell \textit{FCMLI} has $2N$ switches in each leg, and  it produces $(N+1)$ voltage levels from $\frac{-V_{dc}}{2}$ to $\frac{V_{dc}}{2}$, where $V_{dc}$ denotes the input voltage source.
\parisaa{The voltage of capacitors should be controlled in such a way that each cell experiences the same voltage stress. Thus,}  
the average voltage of each capacitor should be kept fixed at multiples of $\frac{V_{dc}}{N}$. The nearest capacitor to the load has $\frac{V_{dc}}{N}$, whilst the capacitor that is connected to the voltage source has $V_{dc}$. \parisaa{Therefore, each blocking switch will sustain a voltage equal to $\frac{V_{dc}}{N}$, where $N$ is the number of cells.}  \parisaa{Fig. \ref{fig: Topology_system} shows a three-phase four-level flying  capacitor \textit{(FC)} inverter with three cells. } The capacitor voltage ratio should be set as $V_{1x}:V_{2x}:V_{3x} = 1:2:3$ with $x\in \{a,b,c\}$ for generating the four-level voltage waveform. \parisaa{ In fact, by selecting the best switching state, some capacitors will charge and some will discharge during each sampling time to achieve the desired voltage ratio. 
}
Each capacitor's voltage can be estimated using the inverter current $i_x$ and switching states in the corresponding leg, as given in (\ref{eq:1}) and  (\ref{eq:2}). \kasim{ As it can be seen in Fig. \mbox{\ref{fig: Topology_system}}, the third capacitor $C_{3x}$ is in parallel with the input voltage source $V_{dc}$; \ihab{thus}, $V_{3x}$ is equal to $V_{dc}$.} In addition, for a resistive-inductive load (i.e., $RL$-load), the three-phase output voltage can be written as in (\ref{eq:3}), where $R$ and $L$ are the load resistance and inductance, respectively. \kasimm{Table~\ref{table: swtching_states} displays the eight possible switching states of a single-phase \textit{FCMLI} and the corresponding output voltage \cite{defay2010direct,meynard2002multicell}.}

\begin{equation}
V_{1x}= V_{1x}(0)+\frac{1}{C_{1x}} \int_{0}^{t}i_{x}(S_{2x}-S_{1x})dt 
\label{eq:1}
\end{equation}
\begin{equation}
V_{2x}= V_{2x}(0)+\frac{1}{C_{2x}} \int_{0}^{t}i_{x}(S_{3x}-S_{2x})dt 
\label{eq:2}
\end{equation}
\begin{equation}
V_{xN}= Ri_{x}+ L \frac{di_{x}}{d_{t}}+ \frac {1}{3}(V_{aN}+V_{bN}+V_{cN})  
\label{eq:3}
\end{equation}

\begin{table}[ht]
\caption{Switching states for a single-phase \textit{FC} converter}
\small\addtolength{\tabcolsep}{-3.5pt} 
\centering
\begin{tabular}{|c||c|c|c||c|c|}
\rowcolor{applegreen}
\hline
$V_i$&$S_{1x}$        & $S_{2x}$         & $S_{3x}$  & Output voltage $V_{xN}$ & Equation \\
\hline\hline
$V_0$&0            & 0              & 0     & $-V_{dc}/2$    & $-V_{dc}/2$ \\ \hline 
$V_1$&1            & 0             & 0      & $-V_{dc}/6$  & $V_{1x}-V_{dc}/2$ \\ \hline 
$V_2$&0            & 1             & 0      & $-V_{dc}/6$     & $V_{2x}-V_{1x}-V_{dc}/2$  \\ \hline
$V_3$&1            & 1             & 0      & $V_{dc}/6$    & $V_{2x}-V_{dc}/2 $\\ \hline  
$V_4$&0            & 0             & 1      & $-V_{dc}/6$     & $V_{dc}/2-V_{2x}$  \\ \hline
$V_5$&1            & 0             & 1      & $V_{dc}/6$   & $V_{dc}/2-V_{2x}+V_{1x}$\\ \hline
$V_6$&0            & 1             & 1      & $V_{dc}/6$    & $V_{dc}/2-V_{1x}$     \\ \hline
$V_7$&1            & 1              & 1       & $V_{dc}/2$    & $V_{dc}/2$ \\ \hline
\end{tabular}
\label{table: swtching_states}
\end{table}

\section{Model Predictive Control for \textit{FCMLI}}\label{MPC}
\begin{figure}[!ht]
\renewcommand{\figurename}{Fig.}
\begin{center}

\begin{tikzpicture}[font=\small,thick, scale=0.85]
 
\node[draw,
    rounded rectangle,
    minimum width=1.5cm,
    minimum height=0.5cm,
    fill=red!30] (block1) {Start};
    
\node[draw,
    rounded corners,
    below= 0.3cm of block1,
    minimum width=1.0cm,
    minimum height=0.5cm,
     fill=blue!10,
] (block2) { $i = 1$ };

\node[draw,
    align=center,
    rounded corners,
    trapezium, 
    trapezium left angle = 65,
    trapezium right angle = 115,
    trapezium stretches,
    below= 0.3cm  of block2,
    minimum width=3.cm,
    minimum height=1.0cm,
    fill=green!10
  ] (block3) {\textcolor{red}{Measured variables at instant $k$}\\
$V^k_{1x}, V^k_{2x}, i^k_{x}$};
 
 \node[draw,
    diamond,
    below= 0.3cm  of block3,
    minimum width=1.7cm,
    fill=orange!30,
    inner sep=0] (block4) { $ i \leq N $};

\node[draw,
    rounded corners,
    align=center,
    below = 0.5cm of block4,
    minimum width=3.3cm,
    minimum height=2.cm,
    fill=blue!10,
    ] (block5) {\textcolor{red}{Predict variables at instant $k+1$}\\
    $V^{k+1}_{1x} = V^k_{1x}+\frac{T_{s}}{C_{1x}} i^k_{x} (S^i_{2x}-S^i_{1x})$ \\ $V^{k+1}_{2x} = V^k_{2x}+\frac{T_{s}}{C_{x}} i^k_{x} (S^i_{3x}-S^i_{2x})$\\
    \hspace*{6pt}$i^{k+1}_{x} = (V_{xN}-V_{ON})M_{1} + i^k_{x} M_{2}$};
    
\node[draw,
    align=center,
    rounded corners,
    below = 0.3cm of block5,
    minimum width=3.0cm,
    minimum height=1.5cm,
    fill=blue!10,
    ] (block6) {\textcolor{red}{Calculate cost function at $k+1$} \\
    $ J_{i}= \lambda_{1}(i^*_{x}-i^{k+1}_{x})^2 +$\\
    $\lambda_{2}(V^*_{1x}-V^{k+1}_{1x})^2+ \lambda_{2}(V^*_{2x}-V^{k+1}_{2x})^2 $ 
    };
    
 \node[draw,
    diamond,
    below= 0.3cm of block6,
    minimum width=1.8cm,
    fill=orange!30,
    inner sep=0] (block7) { $ J_{i} \leq e $};
\node[draw,
    rounded corners,
    right = of block7,
    minimum width=2.2cm,
    minimum height=0.5cm,
    fill=blue!10,
    ] (block8) {$ e = J_{i}  , S_{opt} = S_{i}$};
 
\node[draw,
    rounded corners,
    below =0.5cm of block7,
    minimum width=1.0cm,
    minimum height=0.5cm,
    fill=blue!10,
    ] (block9) {$i^{++} $};

\node[draw,
    rounded rectangle,
    below =0.3cm of block9,
    minimum width=2.4cm,
    minimum height=0.5cm,
    fill=red!30] (block10) {Return $ S_{opt} $};
    
\node [left=2.5 cm of block4] (com) {};
\coordinate [left=2.5cm of block4] (com);

\node [right= 3.8 cm of block4] (com2) {};
\coordinate [right=3.8 cm of block4] (com2);
    
\draw[-latex] (block1) edge (block2)
    (block2) edge (block3)
    (block3) edge (block4);
 
\draw[-latex] (block4) edge node[anchor=east]{\textcolor{blue}{\textit{TRUE}}}(block5);

\draw[-latex, rounded corners=10pt] (block4) -- node[anchor=south] {\hspace*{-2.5cm}\textcolor{red}{\textit{FALSE}}} (com2) |- (block10);
\draw[-latex] (block5) edge (block6);
\draw[-latex] (block6) edge (block7);
\draw[-latex] (block7) -- node[anchor=south] {\textcolor{blue}{\textit{TRUE}}} (block8);
\draw[-latex] (block7) edge node[anchor=east] {\textcolor{red}{\textit{FALSE}}} (block9);
\draw [-latex, rounded corners=10pt] (block9) -| (com) -- (block4);
\draw[-latex, rounded corners=10pt] (block8) |- (block9);

\end{tikzpicture}
 
\caption{Flowchart diagram of \textit{MPC} scheme for \textit{FCMLI}.}
\label{fig: flow}
\end{center}
\end{figure}
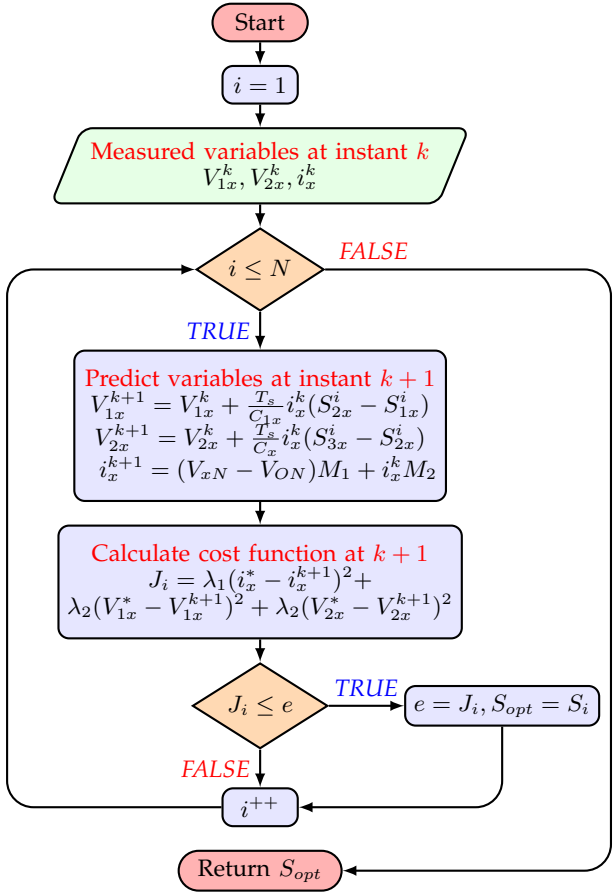

Model predictive control \textit{(MPC)} is one of the well-established and promising model-based control methods that is successfully used in power electronics. The key idea behind \textit{MPC} is the use of the system model for predicting the future behavior of the variables to be controlled, considering a certain time horizon. 
To do so, it needs, in our case, the discrete model of the converter. The optimal control signal, i.e., switching states as described in Table~\ref{table: swtching_states}, that minimizes the cost function is then determined. The entire \textit{MPC} procedure for \textit{FCMLI} is shown in Fig. \ref{fig: flow}, which can be described step-by-step as follows:
\begin{enumerate}
\item At sampling instant $k$, the controlled variables (namely,  $V^k_{1x}, V^k_{2x}, i^k_{x}$) should be measured.
\item Those controlled variables are then predicted at instant $k+1$ based on the discrete model of the converter given in (\ref{eq:4}) and (\ref{eq:5}), where $M_{1}=\frac{T_{s}}{L} $, $M_{2} = 1- \frac{RT_{s}}{L} $, and $T_{s}$ is sampling time; while $V^{k+1}_{1x},V^{k+1}_{2x}, i^{k+1}_{x}$ are the flying capacitor voltages and output current at instant $k+1$. 
\begin{equation}\label{eq:4}
\begin{aligned}
 V^{k+1}_{1x} = V^k_{1x}+\frac{T_{s}}{C_{1x}} i^k_{x} (S^i_{2x}-S^i_{1x})\\
 V^{k+1}_{2x} = V^k_{2x}+\frac{T_{s}}{C_{2x}} i^k_{x} (S^i_{3x}-S^i_{2x})
\end{aligned}
\end{equation}
\begin {equation}
 i^{k+1}_{x} = (V_{xN}-\textcolor{black}{V_{ON}})M_{1} + i^k_{x} M_{2}
\label{eq:5}
\end {equation}
where $V_{ON}$ is the common mode voltage.
\item After defining a proper cost function $J_{i}$, as in (\ref{eq:6}), it should be calculated for the current switching states $S_i$ based on the desired value of each controlled variable, namely, $V^*_{1x}, V^*_{2x}, i^*_{x}$.
\begin{equation}
J_{i}= \lambda_{1}(i^*_{x}-i^{k+1}_{x})^2+\lambda_{2}(V^*_{1x}-V^{k+1}_{1x})^2+ \lambda_{2}(V^*_{2x}-V^{k+1}_{2x})^2  
\label{eq:6}
\end{equation}
\item As the main objective of the optimization problem is to find the optimum switching state ($S_{opt}$) that minimizes the cost function,  the cost function of the current switching state $J_i$ is compared with the smallest previous value $e$.

\item Steps 2) to 4) are repeated for all possible switching states given in Table~\ref{table: swtching_states}.
\item Finally, the optimum switching state ($S_{opt}$) is applied at the next sampling instant.
\end{enumerate}

  In Fig. \ref{fig: flow}, $N$ refers to the total number of the possible switching states in each phase, which is equal to 8 in this study. 

One of the most critical stages of \textit{MPC}, as previously stated, is determining the optimal switching state ($S_{opt}$) that minimizes a given cost function. Different forms of cost functions have been established, in this regard, which are used according to the requirements from the \textit{MPC} design. In this work, the cost function is expressed as (\ref{eq:6}), where $i^*_{x}$ and $i^{k+1}_{x}$ are the reference and estimated values of the output current at $k+1$, respectively. Similarly, $V^{k+1}_{1x}, V^{k+1}_{2x}$ and  $V^*_{1x}, V^*_{2x}$ refer to the first and second capacitor voltages and their desired values at $k+1$, respectively. $\lambda_{1}$ and $\lambda_{2}$ are weighting factors of the cost function that determines the importance of the related term in the cost function. For instance, a higher weighting factor indicates the importance of the variable in determining the optimal switching state, thus it shows the priority of that variable in determining the optimal switching state. 
As previously described, the cost function must be evaluated for all switching states at each sampling time, which can be difficult in systems with many possible switching states such as multilevel inverters; thereby, it needs a strong microprocessor to deal with this high computational burden. In fact, the major drawback of \textit{MPC} is its optimization procedure that should be solved online, which imposes a large amount of \textit{real-time} calculation. This is an important motivation for proposing an \textit{End-to-End} \textit{(E2E)} control strategy, such as \textit{ANN}-Based control strategy, that generates directly the optimal switching states of the inverter, without the need for either: (i) the mathematical model of the inverter and (ii) a pre-defined cost function to be minimized \cite{mohamed2019neural}.

\section{Proposed \textit{ANN}-Based Control Strategy}
\label{proposed-ANN-based}
\ihab{This section provides a brief overview of the artificial neural network (\textit{ANN}) followed by a detailed explanation of the proposed \textit{ANN}-based control strategy for controlling the \textit{FCMLI}. Moreover, it studies the influence of considering different input features on the performance of the \textit{ANN}-based control scheme.}

\subsection{Overview of \textit{ANN}}

Artificial neural network (\textit{ANN}) is a subset of machine learning that can be directly defined as a mathematical relationship between input features and targets; in other words, it is a mathematical model that uses learning algorithms to make intelligent decisions based on historical data. It consists of a set of nodes, so-called neurons, that form layers that are connected and executed in parallel.
\begin{figure}[th!]
\renewcommand{\figurename}{Fig.}
  \centering
  \hspace*{-0.2cm}\resizebox{9cm}{!}{\tikzset{%
every input neuron/.style={circle, draw, fill=green!50, minimum size=1cm},
every output neuron/.style={circle, draw, fill=red!50, minimum size=1cm},
every hidden neuron/.style={circle, draw, fill=blue!50, minimum size=1cm},
neuron missing/.style={ draw=none, scale=4, fill=none, text height=0.333cm, execute at begin node=\color{black}$\vdots$},
neuron2 missing/.style={ draw=none, scale=4, fill=none, text height=0.1cm, execute at begin node=\color{black}$\vdots$},
}

\begin{tikzpicture}[x=1.5cm, y=1.5cm, >=latex]
\foreach \m/\l [count=\y] in {1,2,3,missing,4}
  \node [every input neuron/.try, neuron \m/.try] (input-\m) at (0,2.5-\y) {};
\foreach \m [count=\y] in {1,missing,2}
  \node [every hidden neuron/.try, neuron2 \m/.try ] (hidden-\m) at (2,2.2-\y*1.4) {};
\foreach \m [count=\y] in {1,missing,2}
  \node [every output neuron/.try, neuron2 \m/.try ] (output-\m) at (4,1.8-\y*1.2) {};

\foreach \l [count=\i] in {1,2,3,M}
  \draw [<-] (input-\i) -- ++(-0.9,0)
    node [above, midway] {$x_\l$};

\foreach \l [count=\i] in {1,J}
  {
    \node at (hidden-\i) {$H_\l$};
    \draw [latex-, rotate=90] (hidden-\i) -- ++(0.7,0)
    node [above] {$b_{1\l}$};
  }
  
\foreach \l [count=\i] in {1,N}
  {
  \draw [-latex] (output-\i) -- ++(0.9,0)
    node [above, midway] {$y_\l$};
    \draw [latex-, rotate=90] (output-\i) -- ++(0.7,0)
    node [above] {$b_{2\l}$};
  }
  
\foreach \i in {1,...,4}
  \foreach \j in {1,...,2}
    \draw [->] (input-\i) -- (hidden-\j);
    
\foreach \i in {1}
  \foreach \j [count=\c] in {1,J}
    \draw [-latex] (input-\i) --  node [pos=0.5,fill=cyan!10,inner sep=2pt, rounded corners]{$w_{\i\j}$} (hidden-\c);
\foreach \i in {4}
  \foreach \j [count=\c] in {1,J}
    \draw [-latex] (input-\i) --  node [pos=0.33,fill=cyan!10,inner sep=2pt, rounded corners]{$w_{M\j}$} (hidden-\c);
    
\foreach \i in {1,...,2}
  \foreach \j in {1,...,2}
    \draw [->] (hidden-\i) -- (output-\j);
    
\foreach \i in {1}
  \foreach \j [count=\c] in {1,N}
    \draw [-latex] (hidden-\i) --  node [pos=0.4,fill=cyan!10,inner sep=2pt, rounded corners]{$w_{\i\j}$} (output-\c);
    
\foreach \i in {2}
  \foreach \j [count=\c] in {1,N}
    \draw [-latex] (hidden-\i) --  node [pos=0.33,fill=cyan!10,inner sep=2pt, rounded corners]{$w_{J\j}$} (output-\c);
  
  
\node [align=center, above] at (0,2) {\textcolor{green}{Input $\;$Layer}};
\node [align=center, above] at (2,2) {\textcolor{blue}{Hidden $\;$Layer}};
\node [align=center, above] at (4,2) {\textcolor{red}{Output $\;$Layer}};

\end{tikzpicture}}
   \vspace{-0.1in}
  \caption{A typical architecture of a two-layer \textit{ANN} with $M$ inputs, one hidden layer containing $J$ neurons, and $N$ outputs.}
  \label{fig:ANN-structure}
\end{figure}
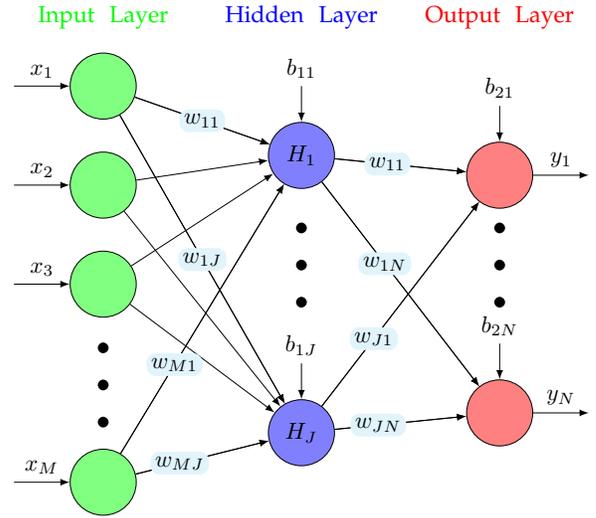
Fig. \ref{fig:ANN-structure} shows the typical architecture of a multi-input multi-output \textit{ANN} that has $M$ input features (i.e., $x_m \; \forall m = \{1, \dots, M\}$) and yields $N$ output targets (namely, $y_n \; \forall n = \{1, \dots, N\}$), where the number of neurons in both input and output layers should be equivalent to the number of data to be processed and the number of targets, respectively.  
Each input feature $x_m$ is multiplied by weighting factors $w_{mj} \; \forall j = \{1, \dots, J\}$ and applied to the hidden layer. In the hidden layer, the bias $b_{1j}$ is added to the sum of the weighted values and an activation function is applied to it in each neuron; then, the results are multiplied by another weighting factors $w_{jm}$ before being passed to the output layer. 
Those correction factors (i.e., biases) allow us to shift the activation functions, which improves the model's fit to the given dataset.
The mathematical formulation of the $n^{\text{th}}$ output can be described as:
\begin{equation}\label{eq:ann-eq}
\begin{aligned}
y_{n}&=h\left(\sum_{j=1}^{J} w_{jn} H_{j}+b_{2n}\right), \quad \forall n = \{1, \dots, N\}, \\
H_{j}&=h\left(\sum_{m=1}^{M} w_{mj} x_{m}+b_{1j}\right), \quad \forall j = \{1, \dots, J\},
\end{aligned}
\end{equation}
where $h$ is the activation function, $H_{j}$ are the outputs of the hidden layer, and $b_{2n}$ refer to the biases of the output layer.
Activation functions $h$ (such as \textit{sigmoid}, \textit{hyperbolic tangent}, \textit{softmax}, and \textit{linear}) are the key part of the neural network design that are used to learn complicated patterns and assess how well the network model learns the training dataset \cite{hornik1991approximation}.
Moreover, there exist different learning methods; in this study, the feed-forward method is used. Unlike recurrent neural networks, connections between different nodes in various layers do not produce a cycle in the feed-forward learning approach.  

\subsection{\textit{ANN}-Based Control Strategy}\label{subsection-ANN-based}
The \textit{ANN}-based control scheme proposed, in this work, undergoes two main phases, namely, training and testing phases.
As mentioned above, the \textit{ANN} should be first trained  \textit{offline}. 
To do so, \textit{MPC} has been used as a baseline control scheme for assessing our proposed control strategy and, in the meanwhile, for collecting data required for training, validation, and testing the proposed neural network.
The acquired \textit{MPC} dataset is used to train the \textit{ANN} to anticipate the optimal switching state $S_{opt}$ at the instant $k+1$. Afterwards, once the proposed neural network is fine-tuned, it can be successfully used \textit{online} for controlling the \textit{FCMLI}, instead of relying on the predictive control strategies.

\begin{figure}[!ht]
\renewcommand{\figurename}{Fig.}
\begin{center}
\definecolor{cyan}{rgb}{0.0, 1.0, 1.0}

\tikzset{every picture/.style={line width=0.75pt}} 

\begin{tikzpicture}[x=0.75pt,y=0.75pt,yscale=-1,xscale=1]

\draw [color={rgb, 255:red, 0; green, 0; blue, 0 }  ,draw opacity=1 ]   (83,99.8) -- (117.57,99.79) ;
\draw [shift={(120.57,99.79)}, rotate = 539.98] [fill={rgb, 255:red, 0; green, 0; blue, 0 }  ,fill opacity=1 ][line width=0.08]  [draw opacity=0] (8.93,-4.29) -- (0,0) -- (8.93,4.29) -- cycle    ;
\draw [color={rgb, 255:red, 0; green, 0; blue, 0 }  ,draw opacity=1 ]   (83,119.8) -- (118.57,119.79) ;
\draw [shift={(121.57,119.79)}, rotate = 539.98] [fill={rgb, 255:red, 0; green, 0; blue, 0 }  ,fill opacity=1 ][line width=0.08]  [draw opacity=0] (8.93,-4.29) -- (0,0) -- (8.93,4.29) -- cycle    ;
\draw  [fill=cyan!15] (122,90.4) .. controls (122,88.74) and (122,87.4) .. (123.57,87.4) -- (219,87.4) .. controls (220.66,87.4) and (222,88.74) .. (222,90.4) -- (222,247) .. controls (222,248.66) and (220.66,250) .. (219,250) -- (123.57,250) .. controls (122,250) and (122,248.66) .. (122,247) -- cycle ;
\draw (174,167.64) node  {\tikzset{%
  every input neuron/.style={circle, draw, fill=green!50, minimum size=0.2cm},
every output neuron/.style={circle, draw, fill=red!50, minimum size=0.1cm},
every hidden neuron/.style={circle, draw, fill=blue!50, minimum size=0.1cm},
neuron missing/.style={ draw=none, scale=2, fill=none, text height=0.333cm, execute at begin node=\color{black}$\vdots$},
neuron2 missing/.style={ draw=none, scale=2, fill=none, text height=0.1cm, execute at begin node=\color{black}$\vdots$},
}

\begin{tikzpicture}[scale=0.9, x=1.2cm, y=1.9cm, >=latex]

\foreach \m/\l [count=\y] in {1,2,3,missing,4}
  \node [every input neuron/.try, neuron \m/.try] (input-\m) at (0,2.5-0.5*\y) {};
\foreach \m [count=\y] in {1,missing,2}
  \node [every hidden neuron/.try, neuron \m/.try ] (hidden-\m) at (0.7,2.4-0.7*\y) {};
\foreach \m [count=\y] in {1,missing,2}
  \node [every output neuron/.try, neuron \m/.try ] (output-\m) at (1.3,2-0.5*\y) {};

\foreach \l [count=\i] in {1,2,3,n}
 \draw [<-] (input-\i) -- ++(-0.4,0);


\foreach \l [count=\i] in {1,n}
  \draw [->] (output-\i) -- ++(0.4,0);

\foreach \i in {1,...,4}
  \foreach \j in {1,...,2}
    \draw [->] (input-\i) -- (hidden-\j);

\foreach \i in {1,...,2}
  \foreach \j in {1,...,2}
    \draw [->] (hidden-\i) -- (output-\j);


\end{tikzpicture}};
\draw [color={rgb, 255:red, 0; green, 0; blue, 0 }  ,draw opacity=1 ]   (223,125.8) -- (245.57,125.79) ;
\draw [shift={(248.57,125.79)}, rotate = 539.97] [fill={rgb, 255:red, 0; green, 0; blue, 0 }  ,fill opacity=1 ][line width=0.08]  [draw opacity=0] (8.93,-4.29) -- (0,0) -- (8.93,4.29) -- cycle    ;
\draw [line width=0.5]  [dash pattern={on 1pt off 4pt}]  (233,160) -- (233,160) -- (233,167) -- (233,195) ;
\draw [color={rgb, 255:red, 0; green, 0; blue, 0 }  ,draw opacity=1 ]   (223,216.8) -- (246.57,216.79) ;
\draw [shift={(249.57,216.79)}, rotate = 539.97] [fill={rgb, 255:red, 0; green, 0; blue, 0 }  ,fill opacity=1 ][line width=0.08]  [draw opacity=0] (8.93,-4.29) -- (0,0) -- (8.93,4.29) -- cycle    ;
\draw  [fill=cyan!15] (250,119) .. controls (250,117.34) and (251.34,116) .. (253,116) -- (269,116) .. controls (270.66,116) and (272,117.34) .. (272,119) -- (272,225) .. controls (272,226.66) and (270.66,228) .. (269,228) -- (253,228) .. controls (251.34,228) and (250,226.66) .. (250,225) -- cycle ;
\draw [dash pattern={on 3.75pt off 3pt on 7.5pt off 1.5pt}, color=red] (107,71) -- (280,71) -- (280,254) -- (107,254) -- cycle ;
\draw [color={rgb, 255:red, 0; green, 0; blue, 0 }  ,draw opacity=1 ]   (223,146.8) -- (246.57,146.79) ;
\draw [shift={(249.57,146.79)}, rotate = 539.97] [fill={rgb, 255:red, 0; green, 0; blue, 0 }  ,fill opacity=1 ][line width=0.08]  [draw opacity=0] (8.93,-4.29) -- (0,0) -- (8.93,4.29) -- cycle    ;
\draw [color={rgb, 255:red, 0; green, 0; blue, 0 }  ,draw opacity=1 ]   (83,140.8) -- (117.57,140.79) ;
\draw [shift={(120.57,140.79)}, rotate = 539.98] [fill={rgb, 255:red, 0; green, 0; blue, 0 }  ,fill opacity=1 ][line width=0.08]  [draw opacity=0] (8.93,-4.29) -- (0,0) -- (8.93,4.29) -- cycle    ;
\draw [color={rgb, 255:red, 0; green, 0; blue, 0 }  ,draw opacity=1 ]   (83,160.8) -- (118.57,160.79) ;
\draw [shift={(121.57,160.79)}, rotate = 539.98] [fill={rgb, 255:red, 0; green, 0; blue, 0 }  ,fill opacity=1 ][line width=0.08]  [draw opacity=0] (8.93,-4.29) -- (0,0) -- (8.93,4.29) -- cycle    ;
\draw [color={rgb, 255:red, 0; green, 0; blue, 0 }  ,draw opacity=1 ]   (82,180.8) -- (116.57,180.79) ;
\draw [shift={(119.57,180.79)}, rotate = 539.98] [fill={rgb, 255:red, 0; green, 0; blue, 0 }  ,fill opacity=1 ][line width=0.08]  [draw opacity=0] (8.93,-4.29) -- (0,0) -- (8.93,4.29) -- cycle    ;
\draw [color={rgb, 255:red, 0; green, 0; blue, 0 }  ,draw opacity=1 ]   (82,200.8) -- (117.57,200.79) ;
\draw [shift={(120.57,200.79)}, rotate = 539.98] [fill={rgb, 255:red, 0; green, 0; blue, 0 }  ,fill opacity=1 ][line width=0.08]  [draw opacity=0] (8.93,-4.29) -- (0,0) -- (8.93,4.29) -- cycle    ;
\draw [color={rgb, 255:red, 0; green, 0; blue, 0 }  ,draw opacity=1 ]   (83,219.8) -- (117.57,219.79) ;
\draw [shift={(120.57,219.79)}, rotate = 539.98] [fill={rgb, 255:red, 0; green, 0; blue, 0 }  ,fill opacity=1 ][line width=0.08]  [draw opacity=0] (8.93,-4.29) -- (0,0) -- (8.93,4.29) -- cycle    ;
\draw [color={rgb, 255:red, 0; green, 0; blue, 0 }  ,draw opacity=1 ]   (83,239.8) -- (118.57,239.79) ;
\draw [shift={(121.57,239.79)}, rotate = 539.98] [fill={rgb, 255:red, 0; green, 0; blue, 0 }  ,fill opacity=1 ][line width=0.08]  [draw opacity=0] (8.93,-4.29) -- (0,0) -- (8.93,4.29) -- cycle    ;
\draw [fill=blue!15]  (292.54,145.95) .. controls (292.54,144.29) and (293.89,142.95) .. (295.54,142.95) -- (346,142.95) .. controls (347.66,142.95) and (349,144.29) .. (349,145.95) -- (349,193.57) .. controls (349,195.23) and (347.66,196.57) .. (346,196.57) -- (295.54,196.57) .. controls (293.89,196.57) and (292.54,195.23) .. (292.54,193.57) -- cycle ;
\draw    (348.19,143.66) -- (293.35,195.86) ;
\draw [line width=1.5]    (305.45,153.52) -- (319.96,153.52) ;
\draw [line width=1.5]    (305.45,159.24) -- (319.96,159.24) ;
\draw  [line width=1.5]  (319.96,182.11) .. controls (323.57,177.85) and (325.28,177.83) .. (328.84,182.11) .. controls (332.4,186.38) and (334.06,186.41) .. (337.71,182.11) ;
\draw [color={rgb, 255:red, 0; green, 0; blue, 0 }  ,draw opacity=1 ]   (271.57,187.48) -- (289.2,187.47) ;
\draw [shift={(292.2,187.47)}, rotate = 539.97] [fill={rgb, 255:red, 0; green, 0; blue, 0 }  ,fill opacity=1 ][line width=0.08]  [draw opacity=0] (8.93,-4.29) -- (0,0) -- (8.93,4.29) -- cycle    ;
\draw [color={rgb, 255:red, 0; green, 0; blue, 0 }  ,draw opacity=1 ]   (272.38,152.45) -- (290,152.44) ;
\draw [shift={(293,152.44)}, rotate = 539.97] [fill={rgb, 255:red, 0; green, 0; blue, 0 }  ,fill opacity=1 ][line width=0.08]  [draw opacity=0] (8.93,-4.29) -- (0,0) -- (8.93,4.29) -- cycle    ;

\draw (58,83.28) node [anchor=north west][inner sep=0.75pt]   [align=left] {$V_{1x}(k)$};
\draw (58,103.8) node [anchor=north west][inner sep=0.75pt]   [align=left] {$V_{2x}(k)$};
\draw (253,220) node [anchor=north west][inner sep=0.75pt]  [rotate=-270] [align=left] {Apply $S_{opt}(k+1)$};
\draw (224,112) node [anchor=north west][inner sep=0.75pt]   [align=left] {$\text{V}_0$};
\draw (224,202) node [anchor=north west][inner sep=0.75pt]   [align=left] {$\text{V}_7$};
\draw (110,72) node [anchor=north west][inner sep=0.75pt]   [align=left] {\small \textcolor{red}{\textit{ANN}-Based Control Strategy}};
\draw (224,133) node [anchor=north west][inner sep=0.75pt]   [align=left] {$\text{V}_1$};
\draw (58,124.28) node [anchor=north west][inner sep=0.75pt]   [align=left] {$i_{ref}(k)$};
\draw (58,143.8) node [anchor=north west][inner sep=0.75pt]   [align=left] {$i_{x}(k)$};
\draw (55,164.28) node [anchor=north west][inner sep=0.75pt]   [align=left] {$\Delta V_{1x}(k)$};
\draw (58,184.28) node [anchor=north west][inner sep=0.75pt]   [align=left] {$\Delta V_{2x}(k)$};
\draw (58,203.28) node [anchor=north west][inner sep=0.75pt]   [align=left] {$\Delta i_{x}(k)$};
\draw (58,223.28) node [anchor=north west][inner sep=0.75pt]   [align=left] {$S_{opt}(k)$};
\draw (300,128) node [anchor=north west][inner sep=0.75pt] [align=left]
{\small \textcolor{blue}{\textit{FCMLI}}};

\end{tikzpicture}
\caption{Block diagram of the \textit{ANN}-based control strategy, considering $\mathbf{X}_2 = \{V_{1x}, V_{2x},i_{ref}, i_{x}, \Delta V_{1x}, \Delta V_{2x}, 2\Delta i_{x}, S_{opt}\}$ as input features.}
\label{fig: Proposed_model}
\end{center}
\end{figure}
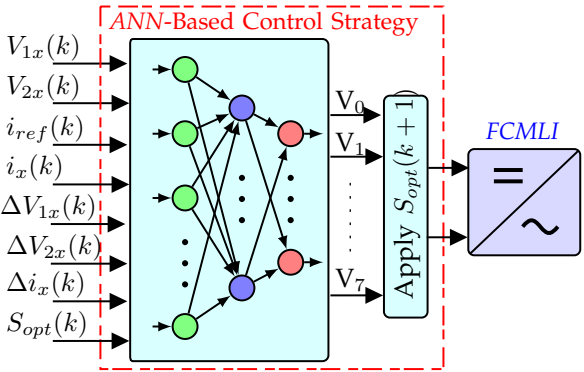

In this study, the input features of our proposed  \textit{ANN}-based control scheme consist of a set of measured variables of the inverter \ihab{that will be discussed in Section~\ref{Input Features Selection and Training Process}}, while the output includes the optimal switching states $S_{opt}$ that should be applied to the inverter at next sampling instant $k+1$, as illustrated in Fig.~\ref{fig: Proposed_model}. More precisely, the output includes a vector of eight elements that identifies the eight voltage vectors ($V_i \; \forall i \in \{0, \dots, 7\}$) (see Table~\ref{table: swtching_states}). Each sampling instant, the index of the \textit{optimum} voltage vector will be only active with a value of one, whereas others will be equal to zero.
\begin{table*}[!ht]
\small\addtolength{\tabcolsep}{-1pt} 
\setlength\extrarowheight{1pt}
\caption{\textit{THD} of the output current $i_{x}$ for different sets of input features, where the lowest \textit{THD} is highlighted in \textcolor{blue}{blue}.}
\centering
\begin{tabular}{|c||c|c|c|c|c|c|c|c|c|c|c|c|c|c|c|c|}
\hline
\rowcolor{applegreen}
Sample No. & $S_{1}$ & $S_{2}$  &$S_{3}$ &$S_{4}$ &$S_{5}$ &$S_{6}$ &$S_{7}$  &$S_{8}$ &$S_{9}$ &$S_{10}$ &$S_{11}$ &$S_{12}$ &$S_{13}$ &$S_{14}$ &$S_{15}$ &$S_{16}$ \\ \hline\hline
$V_{dc} \SI{}{\;[\volt]}$  & 360 & 360  &360     & 342  &378 &360 &360  &350  &340  &360   &360 &340 &350 &360 &350 &350  \\ \hline
$T_{s}\SI{}{\;[\micro\second]}$  & 30 &30 &30 &20 &20 &45 &45  &50 &50 &50  &20 &25  &20 &20 &25 &40\\ \hline
$R\SI{}{\;[\ohm]}$  & 10  &15 &25 &7.5 &15 &15 &8 &9 &11 &10   &10 &10 &12 &7 &9 &9  \\ \hline
$L \SI{}{\;[\milli\henry]}$  &5  &10 &12 &8 &4.5 &9 &10 &9.5 &5  &5.5 &5 &5 &7 &5 &5 &5\\ \hline
$i_{ref}\SI{}{\;[\ampere]}$  &17  &12 &5 &12 &10 &4 &5 &6  &6  &4.35  &12 &10 &8 &15 &12 &12  \\ 
\hline \hline
  \rowcolor{orange!30}
\multicolumn{17}{|c|}{$\mathbf{X_1} = \{V_{1x}, V_{2x}, i_{ref}, i_{x}, \Delta V_{1x}, \Delta V_{2x}, \Delta i_{x}, S_{opt}, V_{ph}\}$} \\ 
\hline
 \textit{THD} [$\%$] & 1.89  &1.29 &3.09 &1.30 &2.02 &4.54 & \textcolor{blue}{\textbf{3.51}} &3.22   &6.48  &8.88 &1.44 &2.01 &1.85 &1.40 &1.73 &2.68 \\ \hline \hline
 \rowcolor{orange!30}
\multicolumn{17}{|c|}{\textcolor{red}{$\mathbf{X_2} = \{V_{1x}, V_{2x}, i_{ref}, i_{x}, \Delta V_{1x}, \Delta V_{2x}, 2\Delta i_{x}, S_{opt}\}$}} \\ \hline 
\textit{THD} [$\%$] & 1.65  &\textcolor{blue}{\textbf{1.06}} &\textcolor{blue}{\textbf{2.36}} & \textcolor{blue}{\textbf{0.93}} &2.07 &\textcolor{blue}{\textbf{4.13}} &3.73 &\textcolor{blue}{\textbf{3.17}}  & \textcolor{blue}{\textbf{6.22}}  &8.01 & \textcolor{blue}{\textbf{1.41}} &2.21 &\textcolor{blue}{\textbf{1.25}} &\textcolor{blue}{\textbf{1.20}} & \textcolor{blue}{\textbf{1.69}} & 3.06 \\ \hline
 \rowcolor{orange!30}\multicolumn{17}{|c|}{\textcolor{black}{$\mathbf{X_3} = \{V_{1x}, V_{2x}, i_{ref}, i_{x}, \Delta V_{1x}, \Delta V_{2x}, \Delta i_{x}, S_{opt}\}$}} \\ \hline 
\textit{THD} [$\%$] & 1.90  &1.32 &3.28 &1.36 & \textcolor{blue}{\textbf{1.94}} &4.38 &3.78 &3.37  &6.29  &8.11 &1.47 &\textcolor{blue}{\textbf{1.98}} &1.43 &1.22 &1.74 &2.69 \\ \hline
\rowcolor{orange!30}
\multicolumn{17}{|c|}{$\mathbf{X_4} = \{V_{2x}, i_{ref}, i_{x}, \Delta V_{1x}, \Delta V_{2x}, \Delta i_{x}, S_{opt}\}$} \\ \hline 
 \textit{THD} [$\%$] & \textcolor{blue}{\textbf{1.59}}  &1.31 &3.64 &1.08 &2.08 &4.39 &3.53 &3.22   &6.57 & \textcolor{blue}{\textbf{7.98}} &1.43 &2.03 &1.71 &1.74 &1.69 &\textcolor{blue}{\textbf{2.67}} \\ \hline \hline
 \rowcolor{orange!30}
\multicolumn{17}{|c|}{$\mathbf{X_5} = \{i_{ref}, i_{x}, \Delta V_{1x}, \Delta V_{2x}, \Delta i_{x}, S_{opt}\}$} \\ \hline
\textit{THD} [$\%$] & 2.69  &1.45 &3.53 &1.32 &2.04 &4.55 &3.71 &3.50 &6.62 &8.11 &1.81 & 2.11 &1.65 &1.76 &1.99 &2.94  \\ \hline
\end{tabular}
\label{table:feature_set}
\end{table*}

\subsection{Input Features Selection and Training Process}\label{Input Features Selection and Training Process}
Basically, selecting the best set of input features is a significant part of designing the \textit{ANN}-based  control strategy. Based on the type of inverters and the main purpose of the study, different features can become efficient in both the training phase and testing phase that assess the capability of learning the mathematical model of the system to be controlled and its dynamics. 
\kasim{\ihab{For the \textit{FCMLI}}, the input features are considered to be a combination of measured variables such as capacitor's voltage, output current and voltage, switching states, and current errors.} 
We have \ihab{empirically} observed that in order to seek the best mapping from the raw input data to the desired outputs, different combinations of input features should be first considered and then assessed. \ihab{Furthermore, it is observed that adding the optimal switching state $S_{opt}$ at instant $k$ to the input features improves the performance of the learning-based control strategy.}

\ihab{To be more specific, five different sets of input features, named: $\mathbf{X}_1,$ $\mathbf{X}_2,$ $\mathbf{X}_3,$ $\mathbf{X}_4$, and $\mathbf{X}_5$, are chosen and then assessed, as illustrated in Table~\ref{table:feature_set}. 
It can be seen, for instance, that $\mathbf{X_1}$ is composed of 9 measurable variables, namely: $\{V_{1x},$ $V_{2x},$ $i_{ref},$ $i_{x},$ $\Delta V_{1x},$ $\Delta V_{2x},$ $\Delta i_{x},$ $S_{opt},$ $V_{ph}\}$, where $\Delta V_{1x} = (\frac{V_{dc}}{3} - V_{1x}),$ $\Delta V_{2x} = (\frac{2V_{dc}}{3}-V_{2x}),$ $\Delta i_{x} = (i_{ref}-i_{x})$, $S_{opt}$ is the optimal switching state at time instant $k$, and  $V_{ph}$ refers to the four-level output voltage. 
To evaluate each input set $\mathbf{X}_i \; \forall i\in \{1,2,...,5\}$, a set of training samples (i.e., $S_{1},S_{2},\dots,S_{16}$) that represent different operating conditions are considered. Each sample has different system parameters such as input voltage $V_{dc},$ sampling time $T_s$, $RL$ impedance, and reference output current $i_{ref}$ (see Table~\ref{table:feature_set}).
Afterward, these samples that have been acquired by \textit{MPC} provide a new dataset, which is then used to train the \textit{ANN} for each $\mathbf{X}_i$.
In the following, the inverter is directly controlled using trained \textit{ANN} with each of the $\mathbf{X}_i$ and the THD of the output current $i_{x}$ is computed. This procedure repeats for each $\mathbf{X}_i$ and the results of different operating conditions (i.e., for sample $S_i \; \forall i\in \{1,2,...,16\}$)  are reported in Table~\ref{table:feature_set}.
In fact, the \textit{THD} of the output current $i_{x}$ is used as a decision parameter to find the best set of input features $\mathbf{X}_i$ to be used in the next sections.
}
We can observe from the preliminary results given in Table~\ref{table:feature_set} that the performance of the proposed control strategy on the basis of $\mathbf{X}_5$, in which the lowest set of features is used, is the worst as the \textit{THD} is quite higher. Furthermore, it can be observed that the four-level output voltage $V_{ph}$ that has been added to $\mathbf{X}_1$, does not have much improvement.
\ihab{For that reason, $\mathbf{X}_2$ is selected as input features for the proposed control scheme given in Fig.~\ref{fig: Proposed_model}, due to its significant improvement compared to other considered sets. $\mathbf{X}_2$ is composed of 8 measurable variables, namely:  $\{V_{1x}, V_{2x},i_{ref}, i_{x}, \Delta V_{1x}, \Delta V_{2x}, 2\Delta i_{x}, S_{opt}\}$.}

Once the effective set of features is selected, \textit{MPC} is again utilized for providing the required dataset for training \ihab{the proposed control scheme given in Fig.~\ref{fig: Proposed_model}}.
\ihab{Similar to Table~\ref{table:feature_set}, various training samples (namely, $S_{1},S_{2},\dots,S_{11}$) have been acquired by \textit{MPC}, each comprising different system parameters that were previously defined in Table~\ref{table:feature_set} (namely, 
$V_{dc}, R, L, T_s, i_{ref}$) in addition to different values of the cell capacitance $C$.
}
\ihab{
The eleven different samples are defined as a percentage of the nominal values of the converter parameters which are as follows: 
$V_{dc} = \SI{360}{\;\volt}$, $C$ = 680 $\mu$F, $L = \SI{10}{\;\milli\henry}$, $R =\SI{15}{\;\ohm}$,  $i_{ref}=\SI{15}{\;\ampere}$, while the \textit{MPC}'s sampling time $T_s$ is indicated in the last column.
For instance, $S_1$ is applied to the \textit{FC} inverter and acquired, considering 
$V_{dc} = \SI{342}{\;\volt}$ (i.e., 95\% of its nominal value), $C$ = 646 $\mu$F, $L = \SI{10}{\;\milli\henry}$, $R =\SI{12}{\;\ohm}$,  $i_{ref}=\SI{14.39}{\;\ampere},$ and $T_s = \SI{30}{\micro\second}$. 
}

\begin{figure}[!ht]
\renewcommand{\figurename}{Fig.}
\begin{center}
\input{FIG/Trainconf.tikz}
\caption{\textcolor{black}{Training} confusion matrix \textcolor{black}{for $70\%$ a training set with a misclassification
error of $10.2\%$, where the number of correctly and incorrectly classified observations are shown in green and red, respectively.}}
\label{fig: validationconf}
\end{center}
\end{figure}

In this work, the total dataset is randomly divided into three sets: training set, validation set, and testing set with $70\%$, $15\%$, and $15\%$ of the total dataset, respectively. The optimum number of hidden layers is determined by varying the number of hidden layers and observing the testing accuracy. In addition, the $trainscg$ function is used as a training function. After the training process, the trained \textit{ANN} can be used to control the four-level three-phase flying capacitor inverter. The \textcolor {black}{training} confusion matrix that assesses the performance of the training is shown in Fig. \ref{fig: validationconf}, where the lowest validation error is $0.1016$ taken from epoch $701$ and the overall accuracy is $89.80\%$ with the total instances numbers of $509,788$. 
The training results are summarized in Table~\ref{table: training results}, demonstrating that the training dataset based on $\mathbf{X}_2$ is properly selected, resulting in an efficient control system.






\begin{table}[!ht]
\caption{Training samples of the \textit{ANN}-based control strategy.}
  \renewcommand\cellrotangle{20}
  \settowidth{\rotheadsize}{HIGH}
  \addtolength{\tabcolsep}{-3pt}
  \centering
  \begin{tabular}{*{7}{l}@{}}
    \addlinespace[2ex]
    \rothead{\rlap{\ihab{Sample} No.}}
     & \rothead{\rlap{$V_{dc} = \SI{360}{\;[\volt]}$}} & \rothead{\rlap{$C =\SI{680}{\;[\micro\farad]}$}} & \rothead{\rlap{$L = \SI{10}{\;[\milli\henry]}$ }} & \rothead{\rlap{$R =\SI{15}{\;[\ohm]}$}}
     &\rothead{\rlap{$i_{ref}=\SI{15}{\;[\ampere]}$}}
      &\rothead{\rlap{ $T_{s} \SI{}{\;[\micro\second]}$}}\\
    \hline
    \hline
    \thead{$S_{1}$} & 0.95 & 0.95  & 1.00     & 0.80    &0.95   & $30$   \\ 
    \thead{$S_{2}$}  & 0.90  & 0.85  & 0.95  & 0.70    &0.90    & $10$   \\ \thead{$S_{3}$} &1.25  & 0.90   & 1.20   &1.10     & 1.15  & $50$   \\ 
    \thead{$S_{4}$} &1.10   & 1.05  &1.50    &1.23    &1.05   &$60$  \\ 
    \thead{$S_{5}$}  &1.00     &1.10    & 1.05  &1.40     &0.75   &$15$ \\
    \thead{$S_{6}$} &1.00     & 0.98  &0.75   &1.30     &0.65   &$18$    \\  \thead{$S_{7}$} &1.15   & 1.20   &0.80    &1.17    &0.55   &$50$    \\
    \thead{$S_{8}$} &1.00      &1.07   & 0.88  &0.77    &0.85   &$30$    \\
    \thead{$S_{9}$} &1.00      &1.00      & 0.98  &0.87    &0.50    &$40$  \\
    \thead{$S_{10}$} &1.00     &1.00      & 0.90   &0.10     &2.00      &$5$ \\
    \thead{$S_{11}$} &1.00     &1.00      & 0.90   &0.10     &2.00      &$15$  \\ \hline
  \end{tabular}
  \label{table: MPC_conditions}
\end{table}
\begin{table}[!ht]
\caption{Training results.}
\centering
\begin{tabular}{l c c} 
\hline
 No. instances & Accuracy & Error (Epoch) \\
 \hline\hline  
$509,788$  &$89.80 \%$ &$0.1016\, (701)$\\
\hline
\end{tabular}
\label{table: training results}
\end{table}
\begin{figure}[!ht]
\renewcommand{\figurename}{Fig.}
\begin{center}
\includegraphics[scale=0.95]{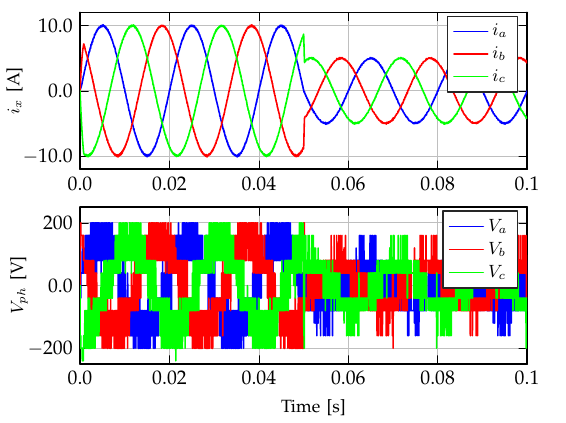}
\vspace*{-17pt}
\caption{Dynamic response of \textit{ANN}-based control strategy: Output current and voltage when the reference current $i_{ref}$ changes from $i_{ref}=\SI{10}{\ampere}$  to  $i_{ref}=\SI{5}{\ampere}$ at $t=\SI{0.05}{\second}$.}
\label{fig: output_daynamic ann-based current}
\end{center}
\end{figure}
\begin{figure}[!ht]
\renewcommand{\figurename}{Fig.}
\begin{center}
\includegraphics[scale=0.95]{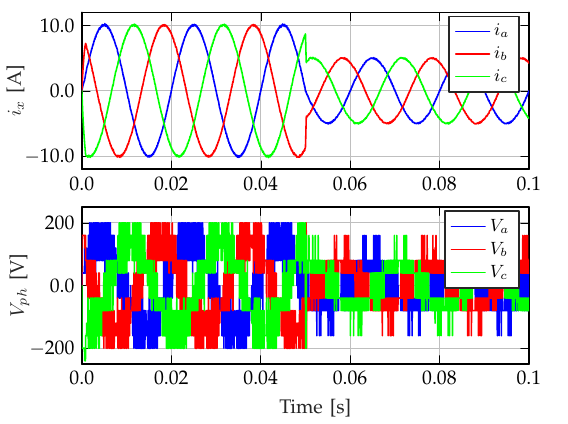}
\vspace*{-17pt}
\caption{Dynamic response of \textit{MPC}: output current and voltage when the reference current $i_{ref}$ changes from $i_{ref}=\SI{10}{\ampere}$ to $i_{ref}=\SI{5}{\ampere}$ at $t=\SI{0.05}{\second}$.}
\label{fig:output_daynamic-MPC-current}
\end{center}
\end{figure}
\begin{figure}
\renewcommand{\figurename}{Fig.}
\begin{center}
\includegraphics[scale=0.95]{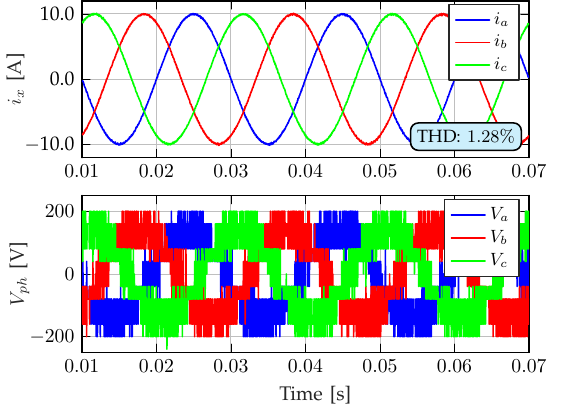}
\vspace*{-17pt}
\caption{Simulation results of \textit{ANN}-based control strategy: Output current and four-level voltage.}
\label{fig: output ann-based current} 
\end{center}
\end{figure}
\begin{figure}%
\renewcommand{\figurename}{Fig.}
\begin{center}
\includegraphics[scale=0.95]{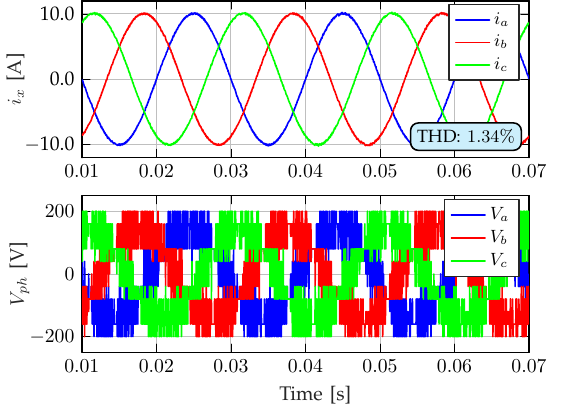}
\vspace*{-17pt}
\caption{Simulation results of \textit{MPC}: Output current and four-level voltage.}
\label{fig: output MPC current}
\end{center}
\end{figure}
\begin{figure}[!ht]
\renewcommand{\figurename}{Fig.}
\begin{center}
\vspace*{-10pt}
\includegraphics[scale=0.8]{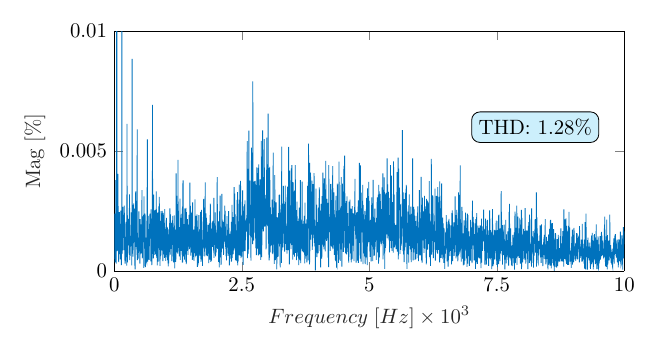}
\vspace*{-22pt}
\caption{Frequency spectrum of output current for the \textit{ANN}-based control strategy.}
\label{fig: ANN_THD}. 
\end{center}
\end{figure}
\begin{figure}[!ht]
\renewcommand{\figurename}{Fig.}
\begin{center}
\includegraphics[scale=0.8]{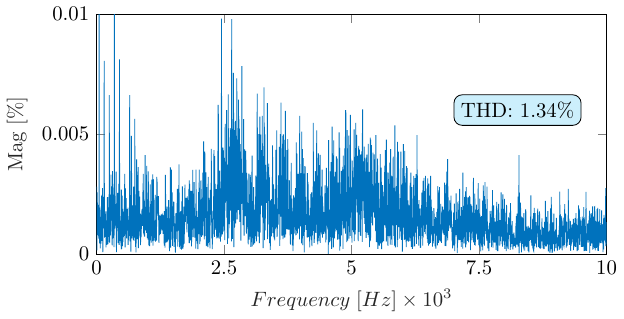}
\vspace*{-9pt}
\caption{Frequency spectrum of output current for the \textit{MPC} scheme.}
\label{fig: MPC_THD}. 
\end{center}
\end{figure}
\begin{figure}[!ht]
\renewcommand{\figurename}{Fig.}
\begin{center}
\includegraphics[scale=0.95]{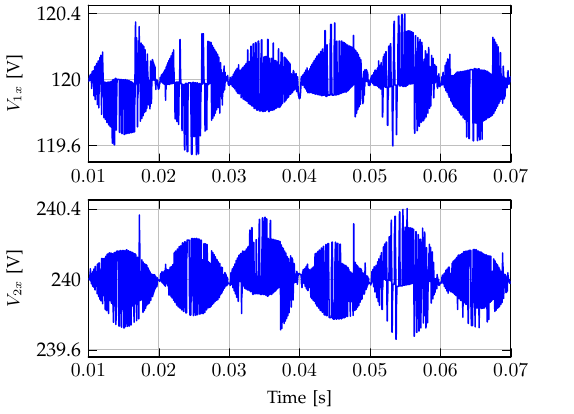}
\vspace*{-15pt}
\caption{Simulation results of \textit{ANN}-based control strategy: Voltages of the capacitors.}
\label{fig: ANN v}
\end{center}
\end{figure}

\section{Simulation Results and Discussion}\label{simulations}
In this section, the \textit{ANN}-based control strategy is simulated using MATLAB/Simulink software and its performance is compared to that of the conventional \textit{MPC}. 
The parameters of the \textit{FCMLI} given in Fig.~\ref{fig: Topology_system} are listed in Table~\ref{table: Inverter Parameters}.
\begin{table}[!ht]
\caption{Parameters of the \textit{FCMLI}.}
\centering
\begin{tabular}{l c} 
\hline
 Parameter & Value \\
 \hline\hline  
$R$  & \SI{15}{\;[\Omega]}\\
$L$  & \SI{10}{\;[\milli\henry]}\\
$C$ & \SI{680}{\;[\micro\farad]}\\
$V_{dc}$ & \SI{360}{\;[\volt]}\\
$T_s$  & \SI{30}{\;[\micro\second]}\\
$f$  & \SI{50}{\;[\hertz]}\\
\hline
\end{tabular}
\label{table: Inverter Parameters}
\end{table}

\begin{figure}[!ht]
\renewcommand{\figurename}{Fig.}
\begin{center}
\includegraphics[scale=0.95]{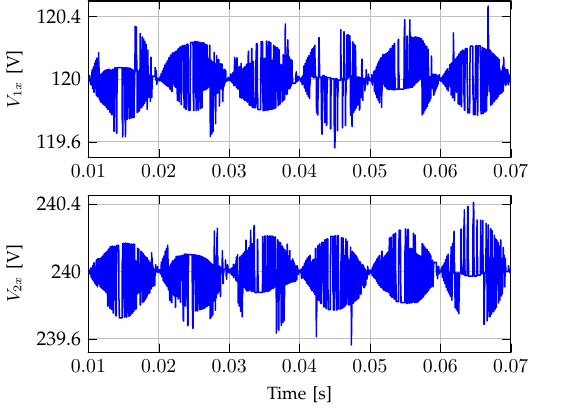}
\vspace*{-15pt}
\caption{Simulation results of \textit{MPC}: Voltages of the capacitors.}
\label{fig: MPC v}
\end{center}
\end{figure}

The dynamic behavior of the \textit{ANN}-based and \textit{MPC} control schemes are illustrated in Figs. \ref{fig: output_daynamic ann-based current} and \ref{fig:output_daynamic-MPC-current}, respectively. Both figures show the \textcolor{black}{output current $i_x$} and four-level output voltage \textcolor{black}{$V_{ph}$} under a sudden change of reference current $i_{ref}$ from \SI{10}{\ampere} to \SI{5}{\ampere} at $t = \SI{0.05}{\second}$. 
It is observed that both methods have similar and good performance under dynamic conditions, since both control strategies track the reference current properly.
The behavior of our proposed \textit{ANN}-based control strategy under steady-state condition is depicted in Fig. \ref{fig: output ann-based current}, while the behavior of the classical \textit{MPC} is shown in Fig. \ref{fig: output MPC current}. 
It can be seen in the figures that the \textcolor{black}{two control strategies generate a high-quality sinusoidal output current $i_x$ with low distortion.
However, the \textit{THD} of the output current in the case of an \textit{ANN}-based control scheme is slightly lower than that in the case of \textit{MPC}, namely, 1.34\% compared to 1.28\% in case of \textit{ANN}-based method.} 
For more information, the frequency spectrum of the two control methods are shown in Figs. \ref{fig: ANN_THD} and \ref{fig: MPC_THD}.
In addition, the voltages of the capacitors (i.e., $V_{1x}, V_{2x}$) obtained by the  two proposed control strategies are depicted in Figs. \ref{fig: ANN v} and \ref{fig: MPC v}, respectively. 
As can be observed, our proposed control scheme has a proper performance to keep the voltage at its desired value, as well as the \textit{MPC} scheme.
\begin{figure}[!ht]
\renewcommand{\figurename}{Fig.}
\begin{center}
\includegraphics[scale=0.95]{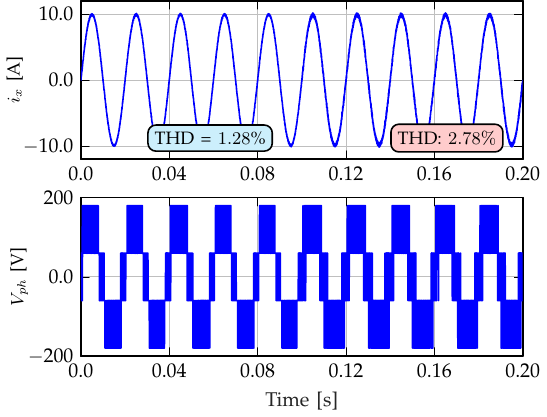}
\vspace*{-14pt}
\caption{Performance of our proposed \textit{ANN}-based control strategy under load inductance variation from $L =\SI{10}{\milli\henry}$ to $L = \SI{5}{\milli\henry}$  at $t=\SI{0.1}{\second}$.}
\label{fig: output ann-based mismatch}
\end{center}
\end{figure}
\begin{figure}%
\renewcommand{\figurename}{Fig.}
\begin{center}
\includegraphics[scale=0.95]{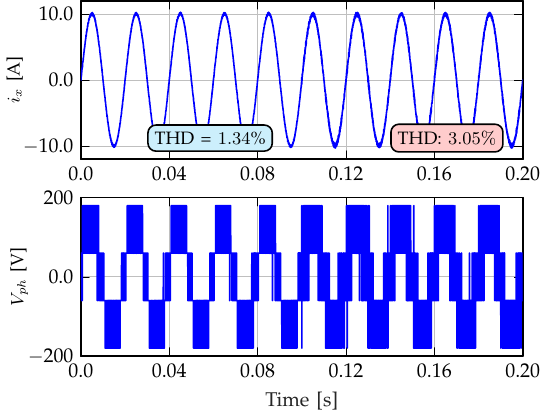}
\vspace*{-14pt}
\caption{Performance of the \textit{MPC} scheme under load inductance variation from $L = \SI{10}{\milli\henry}$ to $L = \SI{5}{\milli\henry}$ at $t=\SI{0.1}{\second}$.}
\label{fig: MPC mismatch} 
\end{center}
\end{figure}
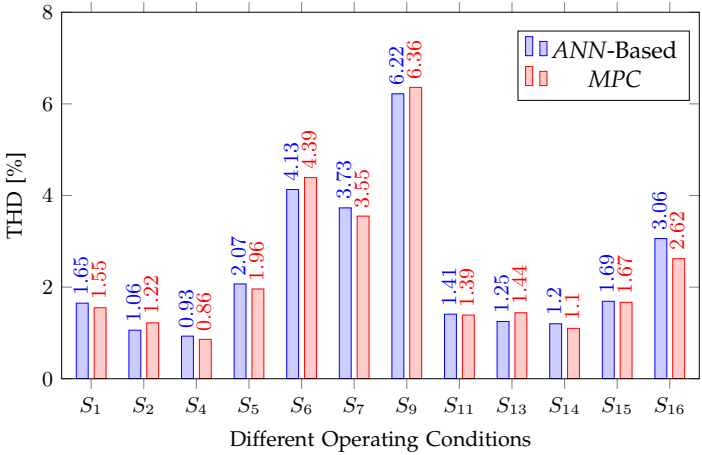
\begin{figure}[!ht]
\renewcommand{\figurename}{Fig.}
\begin{center}
\hspace*{-0.25in}
\begin{tikzpicture}[scale=0.9]  
\begin{axis}[
ybar,height=7.0cm,width=11cm,
enlargelimits=0.0,
legend style={at={(0.85,0.95)}, anchor=north},
ylabel=\small{THD [\%]},
xlabel=\small{Different Operating Conditions},
symbolic x coords={$S_{1}$,$S_{2}$,$S_{4}$,$S_{5}$,$S_{6}$,$S_{7}$,$S_{9}$,$S_{11}$,$S_{13}$,$S_{14}$,$S_{15}$,$S_{16}$},
ybar=2.5pt,
xtick=data,
xticklabel style={font=\small,align=center,rotate=0,},
yticklabel style={font=\small,align=center,
/pgf/number format/.cd,
fixed,
fixed zerofill,
precision=0,
/tikz/.cd
},
ymin=0,
ymax=8,
bar width=5pt,
axis on top,
nodes near coords,
every node near coord/.append style={anchor=mid west, rotate=90},
enlarge x limits={0.051},
]
\addplot[blue,fill=blue!20!white,font=\small,]
coordinates{($S_{1}$,1.65) ($S_{2}$,1.06) ($S_{4}$,0.93) ($S_{5}$,2.07)($S_{6}$,4.13) ($S_{7}$,3.73)($S_{9}$,6.22) ($S_{11}$,1.41)($S_{13}$,1.25)($S_{14}$,1.20)($S_{15}$,1.69)($S_{16}$,3.06)};

\addplot[red,fill=red!20!white,font=\small,]
coordinates{($S_{1}$,1.55) ($S_{2}$,1.22)($S_{4}$,0.86)  ($S_{5}$,1.96) ($S_{6}$,4.39) ($S_{7}$,3.55) ($S_{9}$,6.36) ($S_{11}$,1.39) ($S_{13}$,1.44) ($S_{14}$,1.1)($S_{15}$,1.67)($S_{16}$,2.62)};
\legend{\textit{ANN}-Based,\textit{MPC}}
\end{axis}  
\end{tikzpicture}  
\vspace*{-17pt}
\caption{\textit{THD} comparison of output current between \textit{MPC} and \textit{ANN}-based control strategy.}
\label{fig: THD}
\end{center}
\end{figure}

\hspace*{-5pt}
As previously described, the \textit{ANN}-based control strategy \textcolor{black}{can be used to} deal with the parameter uncertainty issues that arise in real systems.
Thus, to verify the efficiency of the \textit{ANN}-based \textit{MPC} technique in this scope, another simulation study is carried out, where both control strategies are utilized to compare their performance under this condition. 
More precisely, the load inductance $L$ is changed from $\SI{10}{\milli\henry}$ to $\SI{5}{\milli\henry}$ at $t=\SI{0.1}{\second}$; the obtained results are depicted in Figs. \ref{fig: output ann-based mismatch} and \ref{fig: MPC mismatch}. By comparison between these two figures, it can be seen that the proposed \textit{ANN}-based control scheme has better performance in such a situation.  
This can be clearly seen in the \textit{THD} of the output current for the \textit{ANN}-based control scheme that has a 2.78\% compared to 3.05\% in the case of \textit{MPC} after changing the inductance value, which demonstrates that the proposed control scheme is more robust to the parameter uncertainty than the classical \textit{MPC}.

\textcolor{black}{To better assess and show the superiority of the proposed control strategy in learning the mathematical model of the inverter and its dynamics, a comparison of the \textit{THD} of the output current obtained by the two control strategies is represented
in Fig. \ref{fig: THD}, considering the different operating conditions listed in Table~\ref{table:feature_set}.
The obtained results demonstrate that the proposed \textit{ANN}-based control scheme on the basis of the selected input feature (namely, $X_{2}$) has good performance compared to the conventional \textit{MPC}, without requiring a large amount of training data.} 

\begin{figure}[t]
	\centering
	\input{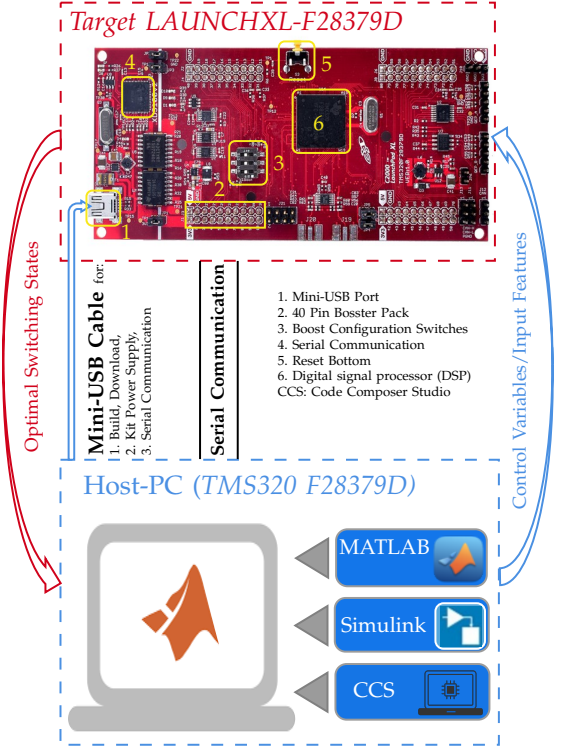}
	\vspace*{-15pt}
	\caption{\parisa{Schematic diagram of the HIL simulation for the proposed system, with the main components and signals flow \cite{zaid2022mpc}.}} 
	\label{fig:HiL-Block}
\end{figure}

\section{\parisaa{HIL Validation and Results}}\label{experimental result}
\intissar{In this section, a rapid prototyping technique was utilized to implement the proposed \textit{ANN}-based
control strategy that previously described in Section~\ref{proposed-ANN-based}. More precisely, a Hardware-in-the-Loop (HIL) simulator was used to verify and validate our proposed control scheme for controlling a three-phase four-level flying capacitor inverter.} 

\intissar{Figure~\ref{fig:HiL-Block} illustrates the basic
components and signal flows of the HIL simulation for the proposed system. 
For a typical HIL simulation, the control strategies (namely, \textit{MPC} and \textit{ANN}-based) are implemented in an external target micro-controller kit (in our study, C2000TM-microcontroller-LaunchPadXL TMS320F28379D kit), while the three-phase four-level flying capacitor inverter and its different loads are simulated and hosted on the personal computer (i.e., Host-PC) as a model in the MATLAB.
The HIL simulation requires cooperation between the Host-PC and the Target LAUNCHXL-F28379D, which is achieved using a virtual serial COM port \cite{ebrahim2021optimal}.
To be more specific, the Host-PC transmits the measured signals (or, input features) such as $V_{1x}, V_{2x},i_{ref}, i_{x}, \Delta V_{1x}, \Delta V_{2x}, \Delta i_{x}$ to the LAUNCHXL-F28379D kit. Afterwards, once the target kit receives these signals, the proposed control strategy selects the optimal switching state $S_{opt}$ to be applied in the next sampling instant. Finally, the Host-PC receives $S_{opt}$ to feed the inverter switches. This procedure will be repeated every sampling time $T_{s}$.
More details about the HIL design procedures, including the simulation setup, Code Composer Studio (CCS), installation, building the target model, and running the Host program, can be found in \cite{mohamed2014implementation}.
}
\begin{figure}[!ht]
\renewcommand{\figurename}{Fig.}
\begin{center}
\includegraphics[scale=0.95]{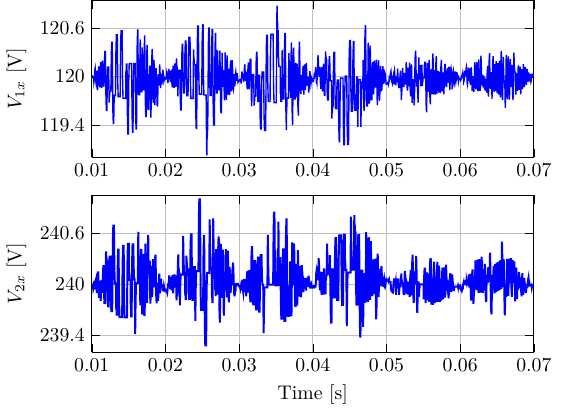}
\vspace*{-17pt}
\caption{\parisaa{HIL results of \textit{ANN}-based control strategy: Voltages of the capacitors.}}
\label{fig: ANN v HIL}
\end{center}
\end{figure}
\begin{figure}
\renewcommand{\figurename}{Fig.}
\begin{center}
\includegraphics[scale=0.95]{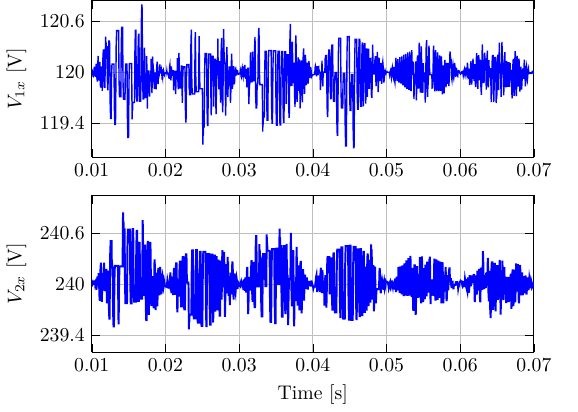}
\vspace*{-17pt}
\caption{\parisaa{HIL results of \textit{MPC} scheme: Voltages of the capacitors.}}
\label{fig: MPC v HIL}
\end{center}
\end{figure}
\begin{figure}
\renewcommand{\figurename}{Fig.}
\begin{center}
\hspace*{-7pt}\includegraphics[scale=0.93]{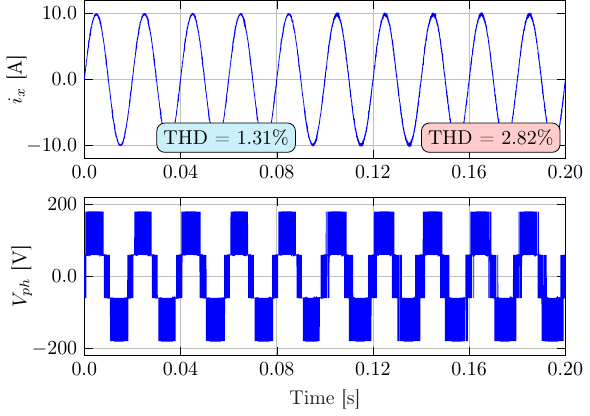}
\vspace*{-17pt}
\caption{\parisaa{HIL results: Performance of the \textit{ANN}-based control scheme under load inductance variation from $L = \SI{10}{\milli\henry}$ to $L = \SI{5}{\milli\henry}$ at $t=\SI{0.1}{\second}$.}}
\label{fig: ANN Mism HIL}
\end{center}
\end{figure}
\begin{figure}
\renewcommand{\figurename}{Fig.}
\begin{center}
\hspace*{-7pt}\includegraphics[scale=0.93]{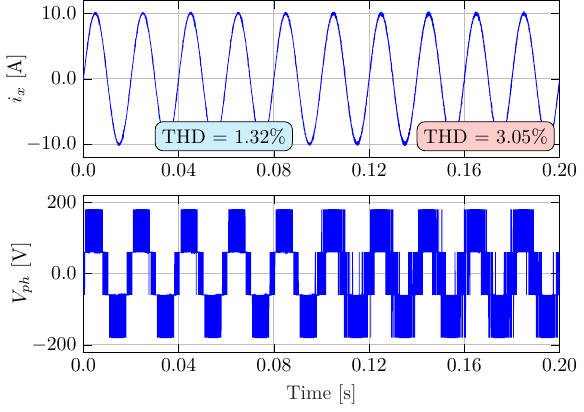}
\vspace*{-17pt}
\caption{\parisaa{HIL results: Performance of the \textit{MPC} scheme under load inductance variation from $L = \SI{10}{\milli\henry}$ to $L = \SI{5}{\milli\henry}$ at $t=\SI{0.1}{\second}$.}}
\label{fig: MPC Mism HIL}
\end{center}
\end{figure}

\begin{figure}[!ht]
\renewcommand{\figurename}{Fig.}
\begin{center}
\hspace*{-0.25in}
\begin{tikzpicture}[scale=0.9]  
\begin{axis}[
ybar,height=8.0cm,width=9.5cm,
enlargelimits=0.0,
legend style={at={(0.8,0.95)}, anchor=north},
ylabel=\small{THD [\%]},
xlabel=\small{Load Power Factor},
symbolic x coords={0.87,0.89, 0.91,0.93},
ybar=2.5pt,
xtick=data,
xticklabel style={font=\small,align=center,rotate=0,},
yticklabel style={font=\small,align=center,
/pgf/number format/.cd,
fixed,
fixed zerofill,
precision=0,
/tikz/.cd
},
ymin=0,
ymax=4,
bar width=15pt,
axis on top,
nodes near coords,
every node near coord/.append style={anchor=mid west, rotate=90},
enlarge x limits={0.15},
]
\addplot[blue,fill=blue!20!white,font=\small,]
coordinates{(0.87,1.85) (0.89,2.11) (0.91,2.23) (0.93,2.48)};

\addplot[red,fill=red!20!white,font=\small,]
coordinates{(0.87,2) (0.89,2.21) (0.91,2.24) (0.93,2.5)};
\legend{Simulation ,HIL}
\end{axis}  
\end{tikzpicture}  
\vspace*{-6pt}
\caption{\kasimm{\textit{THD} comparison of output current between simulation and HIL results at different values of the load's power factor.}}
\label{fig: PF_THD}
\end{center}
\end{figure}
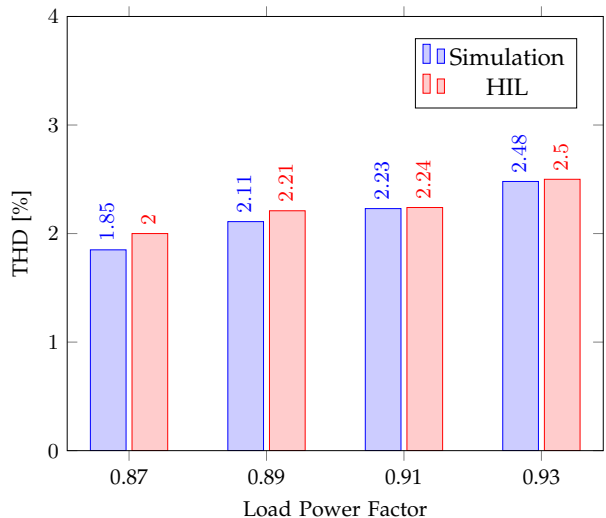

\parisa{After setting up the hardware and the Host-PC as described in \cite{mohamed2014implementation}, the HIL simulation is used to assess the performance of the proposed control schemes in controlling the output current $i_x$, phase voltage $V_{ph}$ and voltage of the capacitors $V_{1x}$ and $V_{2x}$. Figures \ref{fig: ANN v HIL} and \ref{fig: MPC v HIL} show the voltages of the capacitors for each control method. It can be seen that both \textit{ANN}-based and \textit{MPC} have the same and qualified performance to keep the voltage at the desired values, as previously depicted in the simulation section (see Figs. \ref{fig: ANN v} and \ref{fig: MPC v}).
In Figs. \ref{fig: ANN Mism HIL} and \ref{fig: MPC Mism HIL}, we validate the performance of the two control strategies to deal with the parameter uncertainty problem, which is clearly described and tested in the previous section (see Figs. \ref{fig: output ann-based mismatch} and \ref{fig: MPC mismatch}). 
It can be seen in Figs. \ref{fig: ANN Mism HIL} and \ref{fig: MPC Mism HIL} that after changing the inductance from $L = \SI{10}{\milli\henry}$ to $L = \SI{5}{\milli\henry}$ at $t=\SI{0.1}{\second}$, the \textit{ANN}-based control strategy has better performance in comparison with the classical \textit{MPC} scheme.
 \kasimm{This comparison is carried out based on the THD of the output current which is 2.82\% for the proposed control method, after changing the inductance value, compared to 3.05\% when the MPC is employed.}
These results demonstrate the applicability and good performance of our proposed control strategy under realistic conditions.}
\kasimm{To demostatre the efffectiveness of the proposed ANN-based control strategy, the \textit{THD} value of the output current is measured in both simulation and HIL environmnet as shown in Fig. \ref{fig: PF_THD}. The \textit{THD} value is clearly increased as the load power factor decreases due to the reduction of load inductance. The differences between the results of simulation and HIL are negligible, which confirms the proposed \textit{ANN}-based control strategy to be effective in terms of practical applications.}

\section{Conclusions}\label{conclusion}
\textcolor{black}{In this paper, we have proposed an \textit{ANN}-based control strategy for controlling a three-phase four-level flying capacitor inverter with the aim of generating a high-quality sinusoidal output current and improving the robustness against parametric mismatches, especially when compared to conventional \textit{MPC} scheme.}
The proposed control scheme employs the classical \textit{MPC} to generate a reliable dataset of inverter variables to be used as input features by the proposed a feed-forward \textit{ANN}. Afterwards, once the \textit{off-line} training is performed based on the obtained data from \textit{MPC}, it can be used to directly control the inverter without the requirement for a mathematical model of the system. 
Simulation results reveal that \textit{ANN}-based control strategy performs better with respect to the \textit{THD} under most conditions. Additionally, an experiment is carried out for assessing its robustness against parameter uncertainty. 
\parisa{
Moreover, to better assess the performance of the proposed \textit{ANN}-based control method compared to the traditional \textit{MPC}, HIL simulation is utilized to demonstrate its applicability to be run on a DSP kit and prove its efficiency under parameter mismatch problems.}
According to the results, the proposed control scheme showed better performance than the conventional \textit{MPC} under parameter mismatch conditions, owing to the fact that, unlike \textit{MPC}, the proposed technique does not dependent on the system model.


\balance
\bibliographystyle{IEEEtran}
\balance
\bibliography{References}
\balance
\end{document}